\documentclass[useAMS,usenatbib]{mn2e}
\usepackage{graphicx}
\usepackage{epstopdf}
\usepackage{amsmath}
\title[Structure ADAF with Outflows]{Structure of Advection-Dominated Accretion Disks with Outflows: Role of Toroidal Magnetic Field}
\author[Mosallanezhad, Abbassi \& Beiranvand]{A. Mosallanezhad$^{1,2}$\thanks{E-mail: amin@shao.ac.cn (AM);},  S.
Abbassi$^{3,4}$\thanks{E-mail: abbassi@ipm.ir (SA);} and N. Beiranvand$^{5}$\thanks{E-mail: n\_beiranvand@std.du.ac.ir (NB);}\\
$^{1}$Key Laboratory for Research in Galaxies and Cosmology, Shanghai Astronomical Observatory,
Chinese Academy of Sciences,\\ 80 Nandan Road, Shanghai 200030, China\\
$^{2}$University of Chinese Academy of Sciences, 19A Yuquan Road, Beijing 100049, China\\
$^{3}$Department of Physics, School of Sciences, Ferdowsi University of Mashhad, Mashhad, 91775-1436, Iran\\
$^{4}$School of Astronomy, Institute for Research in Fundamental Sciences (IPM), P.O. Box 19395-5531, Tehran, Iran\\
$^{5}$School of Physics, Damghan University, P.O. Box 36715-364, Damghan, Iran}
\begin{document}


\pagerange{\pageref{firstpage}--\pageref{lastpage}} \pubyear{2002}

\maketitle

\label{firstpage}
\begin{abstract}
The main aim of this paper is studying the effect of toroidal magnetic field on the structure of Advection-Dominated Accretion Flows (ADAF) in the presence of the turbulence viscosity and diffusivity due to viscosity and magnetic field respectively. We use self-similar assumption in radial direction to solve the magnetohydrodynamic (MHD) equations for hot accretion disk. We use spherical coordinate $ (r, \theta, \varphi) $ to solve our equation. The toroidal component of magnetic field is considered and all three components of the velocity field $  \mathbf{v}\equiv (v_{r}, v_{\theta}, v_{\varphi}) $ are present in our work. We reduce the equations to a set of differential equations about $ \theta $  and apply the symmetric boundary condition at the equatorial plane of the disk. Our results indicate that the outflow region, where the redial velocity becomes positive in a certain inclination angle $ \theta_{0} $, always exist. The results represent that the stronger the magnetic field, the smaller the inclination angle, $ \theta_{0} $ becomes. It means that a magnetized disk is thinner compared to a non-magnetized disk. According to the work by \citealt{jw}, we can define three regions. The first one is called inflow region, which starts from the disk midplane to a certain inclination $ \theta_0 $ where $ v_{r}(\theta_{0}) = 0 $. In this region, the velocity has a negative value and the accretion material moves toward the central object. The outflow region, where $ v_{r}(\theta) > 0 $, is placed between $ \theta_{0} $ and surface of the disk, $ \theta_{0} < \theta < \theta_{s} $. In this area, the accretion flow moves away from the central object. The third region, which is located between the surface of the disk and the polar axis, is called wind region. This area is very narrow and the material is blown out from the surface in the form of wind. In this paper we consider two parameters to show the magnetic field effects. These parameters include ratio of gas pressure to magnetic pressure in the equatorial plane of the disk, $ \beta_{0} $, and also magnetic diffusivity parameter, $ \eta_{0} $. Numerical calculations of our model have revealed that the toroidal components of magnetic field has a significant effect on the vertical structure of accretion disk.
\end{abstract}

\section{Introduction}

There has been rapid progress in understanding the accretion processes in astrophysics during the past three decades. Today we know that the accretion disks exist in many astrophysical systems such as active galactic nuclei (AGN) and in close X-ray binary systems (XBs) where the accretion disk surrounds a black hole. 

Many theoretical models have been proposed. One of them is the standard accretion disk model presented by \citealt{shakura}. In this model, the disk is assumed to be geometrically thin ($ H/r \ll 1 $), optically thick in the vertical direction and the accreting matter moves with nearly Keplerian velocity. This model explains most of the observational features of X-ray binaries and active galactic nuclei in a highly convincing manner. However, standard disk models cannot reproduce high energy emissions, such as X-ray and gamma rays spectrum.
One of the most important processes that is not considered in the standard accretion disk model is advective cooling. In this model, the accreting matter cools so efficiently that all of the energy released through viscosity can be locally radiated. However, there exist distinct branches of steady disk solutions in which this assumption is violated. In a radiatively inefficient accretion flow (hereafter RIAF), the energy released via viscosity is stored as entropy and transported inward with accretion. So the flow becomes very hot and produces high energy emissions (see \citealt{kato} for a review).

In the case of low mass accretion rate, $ \dot{M} \leq 0.1 L_{E}/c^{2} $, it is expected to have optically thin advection dominated accretion flow (ADAFs/RIAF) with insufficient cooling (\citealt{ichimaru}; \citealt{rees}; \citealt{ni94}, 1995a, 1995b; \citealt{Abramowicz95}). The ADAFs were adopted to the low/hard state of black hole binaries and low-luminous active galactic nuclei (AGNs), such as the supermassive black hole in the center of our galactic (Sgr $ A^{*}) $. If the accretion rate is mildly low $  \dot{M}\leq L_{E}/c^{2} $, the disk is in the Shakura \& Sunyaev standard state (SSD), that the disk is geometrically thin and optically thick. The SSD model can be applied to the high/soft state of  black hole binaries and even for luminous AGNs. Moreover if the mass accretion rate is very high $ \dot{M} \geq L_{E}/c^{2} $, the disk is categorized as a super critical state (slim disk), that the disk is optically and geometrically thick. This bright branch may be applied to the very high state of black hole binaries and super Eddington sources.

The ADAFs models have captured great attention and rapid progress has been made in this field. An extensive discussion on advection dominated accretion disk began by \citealt{ni94} , 1995a, 1995b. They solved the disk structure along $ \theta $ direction using self-similar method in radial direction. Moreover, several numerical simulations (hydrodynamical and magnetohydrodynamical) have been performed to study the structure of ADAFs (see \citealt{stone99}; \citealt{stone01}; \citealt{pen03}; \citealt{de03}; \citealt{okuda05};\citealt{yuan10}; \citealt{mc12}; \citealt{yuan12I} ). This model have been used to interpret the spectra of black hole X-ray binaries in their quiescent or low/hard state as an alternative to the \citealt{sle76} (SLE) solutions. Since ADAFs have large radial velocities, accreting matter carries the thermal energy into the black hole. Thus, the advective energy transport will act against thermal instability. It caused ADAFs have been widely used for explaining the observations of galactic black hole candidates (e.g. \citealt{nmi96}; \citealt{Hameury}), the spectral transition of Cyg X-1 (\citealt{e96}) and multi-wavelength spectral properties of Sgr $ A^{*} $ (\citealt{ni95b}; \citealt{Manmoto}; \citealt{nkh97}; \citealt{YuanQN}).

Many ADAF-like models have been proposed including outflows, convection, etc.  In this regard, the luminous hot accretion flow (LHAF) model by \citealt{yuan01} is a development of ADAF to higher accretion rates. Moreover, the convection-dominated accretion flow (CDAF) model (\citealt{nia00}, \citealt{QG2000}) proposes that accretion flows are convectively unstable. In addition the  adiabatic inflow-outflow solution (ADIOS) model is an additional development that is suggested by \citealt{bb99}, 2004. But recently \citealt{yuan12II} simulations have shown that the convection model is not a complete model to explain outflow.

It has been observationally verified that some of the angular momentum is dissipated outwards in an accretion disk by outflows in the form of wind or jets, from such systems (\citealt{Whelan}; \citealt{Bally}; \citealt{Dionatos09}). These outflows, which characteristically cause the loss of angular momentum, mass and thermal energy, are classified as winds or jets based on their collimation. However, this classification is not exactly clear because of some observed systems both types of outflows (e.g. \citealt{Piran}; \citealt{Blandford}; \citealt{Pudritz}). These kinds of outflows are seen in all sorts of objects from microquasars, YSOs, AGNs etc., which is a good indication of the range of scales involved in this phenomena. The cause of these outflows is the accretion mechanism itself, so it is imperative to understand this mechanism particularly at earlier stages when most of matter is accreted by a central object.

It is found that the rate of the mass loss is proportional to the disk size and mass of the central object possibly by a power-law dependence (\citealt{abassierfan13}). From the first star formation models which considered the accretion flow without wind, it was found that mass accretion is constant $ \sim 2\times10^{-6}M_{\odot}yr^{-1} $, as the collapse proceeds inside-out at local sound speeds. However as a result of outflows, the accretion rate varies with radius as a power-law, with the power-law index of an order of unity (e.g., \citealt{bb99}; \citealt{abassighanbari10}, 2013; \citealt{yuan12I} ). In the earlier phases of star formation, accretion rate is very high ($ \sim 10^{6}M_{\odot}yr^{-1} $) and the mass loss in the system is in the form of winds, with the mass loss rate being $ \sim \times 0.1 \dot{M} $ (\citealt{gh09}). What exactly drives these winds is still not very clear. The emission mechanism of disk winds relies on a variety of physical phenomena such as the effects of magnetic fields, electric fields generated by the relative separation between ions and electrons, electron-positron pairs production and their coupling with the radiation field in the disk winds, etc. (\citealt{Takahara}).


ADAFs solutions predict a high temperature for accretion material. Consequently the gas is ionized which will be affected by the magnetic field. The magnetic field therefore plays an important role in the dynamics of accretion flows and probably in creating of outflow, wind or jet. \citealt{Kaburaki}; \citealt{abassighanbarisalehi}; 2008; 2010; \citealt{ghanbari07}, \citealt{shadmehri}, \citealt{xie2008} and \citealt{bu2009} have tried to solve the magnetohydrodynamics equations of magnetized ADAFs analytically. They showed that the presence of a magnetic field and its associated resistivity can considerably change the picture with regard to accretion flows.

As it was mentioned, Narayan \& Yi used self- similar method in radial direction and solved the disk structure along $ \theta $ direction. The self-similar approach adopted by Narayan \& Yi 1995 is only partially supported by numerical simulations, i.e., there exist a new class of accretion flow, which is hot and optically thin and it is advection-dominated. On the other hand, their shortcoming is also obvious, i.e. the lack of an outflow. Thus, it is not consistent with the new developments in hot accretion flow theory. Also recent theoretical work (Blandford \& Begelman 1999) and numerical simulations (Stone, Pringle \& Begelman 1999; Stone \& Pringle 2001, De Villiers, Hawley, Krolik \& Hirose 2005; Ohsuga \& Mineshige 2011, Yuan et al. 2012a, 2012b) indicates that outflow is commonly observed to be associated with the hot accretion flow. More importantly, there are compelling observational evidences for the existence of outflow in hot accretion systems, e.g. our Galactic Center and NGC 3115 (see Xie \& Yuan 2012 for a short summary). Moreover, numerical simulations indicate that $ v_\theta $ is non-zero (see Stone, Pringle \& Begelman 1999; Ohsuga \& Mineshige 2011; Yuan, Bu \& Wu 2012). Besides, numerical calculations indicate that, it is very difficult to find an outflow solution with $ v_\theta = 0 $ (Narayan \& Yi 1995), whereas with nonzero $ v_\theta $ outflow can be found (e.g. Xue \& Wang 2004) and $ v_r $ will be positive.

Recently \citealt{jw} have solved a set of hydrodynamical equations for accretion disks in the spherical coordinates ($r,\theta,\varphi$) and obtained the explicit structure along the $\theta$ direction. They assumed $v_{\theta}\neq0$ and used a self-similar treatment  that leads to an accretion-outflow solution for ADAFs. The ADAFs solutions with wind were reported previously by \citealt{abassighanbarinajar}, 2010, \citealt{mosallanezhad}  where the effects of wind and outflow are achieved by adding relevant terms in MHD equations. It means that outflow will appear by a power-law assumption in MHD equations.  \citealt{jw} solutions show that when we assume $ v_{\theta}\neq0 $ by adopting proper boundary condition self-similar solution it will lead to constant inflow-outflow behavior. In this manuscript we will try to develop \citealt{jw} solutions by adding a toroidal magnetic field and its correspond resistivity using self-similar method. Recent numerical simulations support that the mass accretion rate, radial velocity and density can be well approximated as power-laws (see, e.g. Stone, Pringle \& Begelman 1999; Yuan, Bu \& Wu 2012). All of these power-law profiles are consistent with the self-similar methodology adopted in our work. Although, the existence of a self-similar solution in no way guarantees that this solution is relevant to real accretion flows (particularly near to boundaries), it is nevertheless likely to provide a good indication of how realistic flows will behave.

The structure of this paper organized as follow. In section 2, we describe basic equations and give a self-similar solution in redial direction corresponding to ADAFs model. Boundary conditions are presented in section 3. In section 4, we show the numerical results and discuss the variation and the physical meaning of each parameter from equatorial plane to the surface of the disk. Conclusions are presented in section 5.

\section{Basic equations and self similar solutions}
\subsection{Basic equations}
In this section we derive the basic equations of optically thin advection dominated accretion flows under magnetohydrodynamics approximation and in a non-relativistic regime. We neglect the effect of self-gravity in our model. The magnetic field is considered with toroidal configurations and in addition to relativistic effects being ignored, we have used Newtonian gravity. The disk is supposed to be turbulent and possess an effective turbulent viscosity. $ \alpha $-prescription for viscosity was adopted. We have assumed that the energy generated due to viscosity and magnetic resistivity is balanced by a combination of radiative and advective cooling. Thus, the resistive MHD equation are involved in continuity equation, equation of motion, energy equation and induction equation that can be respectively written as follows:
\begin{gather}
    \frac{d\rho}{d t} +\rho \mathbf{\nabla} \cdot \mathbf{v} = 0 \label{continuity_main}\\
    \rho \frac{d\mathbf{v}}{dt} = -\rho \mathbf{\nabla}\psi - \mathbf{\nabla}p + \frac{1}{c} \mathbf{J}\times \mathbf{B} + \mathbf{\nabla} \cdot \mathbf{T} \label{motion_main}\\
    \rho \bigg[ \frac{d e}{dt} + p\frac{d}{dt}\Big( \frac{1}{\rho} \Big) \bigg] = Q_{+} - Q_{-}\equiv Q_{adv}  \label{energy_main}\\
    \frac{\partial \mathbf{B}}{\partial t} = \mathbf{\nabla}\times \Big( \mathbf{v}\times \mathbf{B} - \frac{4\pi}{c}\eta \mathbf{J} \Big) \label{induction_main}
\end{gather}
In the above MHD equations, $ \rho $, $ \mathbf{v}\equiv (v_{r}, v_{\theta}, v_{\varphi}) $, $ \psi $, $ p $, $ \mathbf{B} $, $ \mathbf{J} \equiv (c/4\pi) \mathbf{\nabla} \times \mathbf{B} $ and $ \mathbf{T} $ are the mass density, velocity vector, gravitational potential, pressure, magnetic field, current density, and the tensor of viscous stress, respectively.  Here, $ d/dt = \partial/\partial t + \mathbf{v} \cdot \mathbf{\nabla} $. We adopt spherical coordinate $ (r, \theta, \varphi) $ to solve these equation. We consider the Newtonian potential, $ \psi = GM_{*}/r $, where $ G $ is the gravitational constant, $ M_{*} $ is the central object mass and $ r $ is the spherical radial coordinate. In our calculation, we use only the $ r\varphi $-component of the viscous stress tensor, which is $ t_{r \varphi} = \nu \rho \partial (v_{\varphi}/r)/\partial r $, where $ \nu $ is the kinematic viscosity coefficient and will be defined later. In the energy equation, $ e $ is the internal energy that can be expressed as
\begin{equation}
e = \frac{p}{\rho(\gamma - 1)}
\end{equation}
where $ \gamma $ is the ratio of specific heats and is considered as a constant input parameter. On the right hand site of energy equation we have
\begin{equation}
Q_{+} - Q_{-} \equiv Q_{adv}
\end{equation}
here, $ Q_{adv} $ is the advection transfer of energy, $ Q_{-} $ shows the energy loss through radiative cooling and $ Q_{+} $ represent the dissipation rate of heating due to viscosity and resistivity, $ Q_{+} = Q_{vis} + Q_{B} $, which can be defined as
\begin{equation}
Q_{vis} = t_{r \varphi} r \frac{\partial}{\partial r} \big(\frac{v_{\varphi}}{r}\big) = \nu \rho r^{2} \Big( \frac{\partial}{\partial r}\big(\frac{v_{\varphi}}{r}\big) \Big)^{2}
\end{equation}
\begin{equation}
Q_{B} = \frac{\eta}{4 \pi} \Big| \mathbf{\nabla} \times \mathbf{B} \Big|^{2}
\end{equation}
where $ \eta $ is the magnetic diffusivity parameter. $ Q_{B} $ is the hitting rate due to the resistance of fluid against the motion of charges. In other words, $ Q_{B} $ is the Joule hitting or resistivity hitting , that is equal to ${J}\cdot{E} $, where $ E $ is the electric field in the co-moving frame. We consider that $ \nu $ and $ \eta $ have the same units and for both of them use the general case
\begin{gather}
    \nu = \alpha \frac{p_{gas}^{\mu}}{\rho \Omega_{k}}\big(p_{gas} + p_{mag} \big)^{1 - \mu} \equiv \alpha \frac{p}{\rho \Omega_{k}} \Big( 1 + \frac{B^{2}}{8 \pi p}  \Big)^{1 - \mu} \label{nu}\\
    \eta = \eta_{0} \frac{p_{gas}^{\mu}}{\rho \Omega_{k}}\big(p_{gas} + p_{mag} \big)^{1 - \mu} \equiv \eta_{0} \frac{p}{\rho \Omega_{k}} \Big( 1 + \frac{B^{2}}{8 \pi p}  \Big)^{1 - \mu} \label{eta}
\end{gather}
here $ p_{mag}(=B^{2}/8 \pi) $ is the magnetic pressure, $ \Omega_{k} = \sqrt{GM_{*}/r^{3}} $ is the Keplerian angular speed and also $ \alpha, \eta_{0}$ and $ \mu $ are positive constants less than 1. Thus, the right hand side of energy equation will be
\begin{equation}
Q_{adv} \equiv Q_{+} - Q_{-} = Q_{+}\Big(1 - \frac{Q_{-}}{Q_{+}} \Big) = f Q_{+}
\end{equation}
and
\begin{equation}
Q_{+} = \nu \rho r^{2} \Big(\frac{\partial}{\partial r}\big(\frac{v_{\varphi}}{r}\big)\Big)^{2} + \frac{\eta}{4 \pi} \Big| \mathbf{\nabla} \times \mathbf{B} \Big|^{2}
\end{equation}
where $ f $ is the advection parameter which is defined by \citealt{ni94}. Although, this parameter varies with radius $ r $, and depends on the heating and cooling processes, we consider it  as a constant here. We fix $ \mu = 0 $ all over this manuscript. For simplicity, the flow is assumed to be steady and axisymmetic $ (\partial / \partial t = \partial/ \partial \varphi = 0) $. We also consider toroidal component of magnetic field, $ \mathbf{B} = (0, 0,B_{\varphi}) $. Most of the simulations in accretion disks show the toroidal component for magnetic field is enhanced because of rotating disk. We can say utilizing the toroidal component is a right choice in the physics of accretion processes. Now we can reformulate the basic equations (\ref{continuity_main})-(\ref{induction_main}) in spherical coordinates as
\begin{equation}\label{continuty_ref}
\frac{1}{r^{2}}\frac{\partial}{\partial r}(r^{2}\rho v_{r}) + \frac{1}{r \sin\theta}\frac{\partial}{\partial \theta}(\sin\theta \rho v_{\theta}) = 0
\end{equation}
\begin{multline}\label{motionr-ref}
\rho \bigg[ v_{r} \frac{\partial v_{r}}{\partial r}+\frac{v_{\theta}}{r} (\frac{\partial v_{r}}{\partial \theta} - v_{\theta})- \frac{v_{\varphi}^{2}}{r} \bigg] =
- \rho \frac{GM_{*}}{r^{2}} - \frac{\partial p}{\partial r} \\
- \frac{B_{\varphi}}{4 \pi r}\frac{\partial}{\partial r}(r B_{\varphi})
\end{multline}
\begin{multline}\label{motiont_ref}
\rho \bigg[ v_{r} \frac{\partial v_{\theta}}{\partial r} + \frac{v_{\theta}}{r}(\frac{\partial v_{\theta}}{\partial \theta} + v_{r}) - \frac{v_{\varphi}^{2}}{r}\cot \theta \bigg] =
-\frac{1}{r} \frac{\partial p}{\partial \theta} \\
- \frac{1}{4 \pi} \frac{B_{\varphi}}{r \sin \theta} \frac{\partial}{\partial \theta}(B_{\varphi} \sin \theta)
\end{multline}
\begin{multline}\label{motionp_ref}
\rho \bigg[ v_{r} \frac{\partial v_{\varphi}}{\partial r} + \frac{v_{\theta}}{r} \frac{\partial v_{\varphi}}{\partial \theta} + \frac{v_{\varphi}}{r} (v_{r} + v_{\theta} \cot \theta) \bigg] =\\
\frac{1}{ r^{3}} \frac{\partial}{\partial r} \big(\nu \rho r^{4} \frac{\partial}{\partial r}(\frac{v_{\varphi}}{r}) \big)
\end{multline}
\begin{multline}\label{energy_ref}
\frac{\rho}{\gamma - 1} \bigg[ v_{r} \frac{\partial}{\partial r} \big( \frac{p}{\rho} \big) + \frac{v_{\theta}}{r} \frac{\partial}{\partial \theta} \big( \frac{p}{\rho} \big) \bigg] - \frac{p}{\rho} \bigg( v_{r} \frac{\partial \rho}{\partial r} + \frac{v_{\theta}}{r} \frac{\partial \rho}{\partial \theta} \bigg) = \\
f \frac{p}{\rho \Omega_{k}} \big( 1 + \frac{B^{2}}{8 \pi p}  \big) \bigg\{ \alpha \rho r^{2} \big(\frac{\partial}{\partial r}(\frac{v_{\varphi}}{r})\big)^{2} + \frac{\eta_{0}}{4 \pi} \big| \mathbf{\nabla} \times \mathbf{B} \big|^{2} \bigg\}
\end{multline}
\begin{multline}\label{induction_ref}
 \frac{\partial}{\partial r}(r v_{r} B_{\varphi}) + \frac{\partial}{\partial \theta}(v_{\theta} B_{\varphi}) - \frac{\partial}{\partial r}\bigg[ \eta_{0} \frac{p}{\rho \Omega_{k}} \big( 1 + \frac{B^{2}}{8 \pi p}  \big) \frac{\partial}{\partial r}(r B_{\varphi}) \bigg] \\
 - \frac{\partial}{\partial \theta} \bigg[ \frac{p}{\rho \Omega_{k}} \big( 1 + \frac{B^{2}}{8 \pi p}  \big)\frac{\eta_{0}}{r \sin \theta} \frac{\partial}{\partial \theta}(B_{\varphi} \sin \theta) \bigg] = 0
\end{multline}
\subsection{Self Similar Solutions}

The basic equations of the model are a set of partial differential equations which have a very complicated structure. In fact, these partial differential equations are converted into ordinary differential equations by using the assumption of radial self-similarity. Although it must be investigated if there are any critical points or not. Self-similar method is one of the most useful and powerful techniques for solving differential equations. By this method, we can solve a set of coupled differential equations analytically or semi-analytically. The disk equations reduce from partial to ordinary differential equations under the assumption of radial self-similarity, which implies that all quantities are described by power-laws in the spherical radius $ r $ for a fixed inclination angle. This assumption is widely used in the literature of black hole accretion disks (see Narayan \& Yi 1994, 1995, Akizuki \& Fukue 2006, Kato et al. 2008). the self-similar scaling for density and velocities has a very good agreement with recent numerical simulations of accretion disk (Stone, Pringle \& Begelman 1999, De Villiers et al 2005, Beckwith et al. 2008, Yuan et al 2012 ).

Generally self-similar solutions are divided into two chief categories: temporal self-similar answers and spacial self- similar answers. Because the equations of the system depend on time, we can search for answers that describe the temporal change of physical quantities in a way that the change of each quantity at any instant of time is similar to the others. But the second type is the spatially self-similar solutions which are used in this paper. Essentially the spatially self-similar solutions are described the behavior of physical quantities in a manner that at any distance from the center of the system, the difference with the other points are only in a constant factor. As a result, these answers have power functions of positions usually. So we solve a set of equations which depend on $(r, \theta)$. Thus our self-similar solutions must be in the form of $(r, \theta)$. Physical quantities are assumed as unknown powers of $(r, \theta)$. Then we try to put answers in the form of powers in the equations to determine the power in a way that satisfy the equations. If we succeed, it means that equations have self-similar answers.

 We adopt self similar assumption in the radial direction to simplify equation
(\ref{continuty_ref})-(\ref{induction_ref}) as follows:
\begin{gather}
    \rho(r, \theta) = \rho(\theta) r^{-n} \label{self-rho} \\
    v_{r}(r, \theta) = v_{r}(\theta) \sqrt{\frac{GM_{*}}{r}} \label{self-vr} \\
    v_{\theta}(r, \theta) = v_{\theta}(\theta) \sqrt{\frac{GM_{*}}{r}}  \label{self-vtehta} \\
    v_{\varphi}(r, \theta) = v_{\varphi}(\theta) \sqrt{\frac{GM_{*}}{r}} \label{self-vphi} \\
    p(r, \theta) = p(\theta) GM_{*} r^{-n-1} \label{self-p} \\
    B_{\varphi}(r, \theta) = b(\theta) \sqrt{GM_{*}} r^{-\frac{n}{2}-\frac{1}{2}} \label{self-B}
\end{gather}
Our set of self similar solutions is similar to that of Narayan \& Yi 1995 and we add a new equation (\ref{self-B}) for toroidal component of magnetic field. We follow \citealt{xw} and \citealt{jw} to set $ v_{\theta} \neq 0 $ and study the structure of disks with the presence of outflows. With self- similar solutions (\ref{self-rho})-(\ref{self-B}), the equations (\ref{continuty_ref})-(\ref{induction_ref}) are reduced to a set of differential equations as follow
\begin{equation}\label{rho_new}
\rho(\theta) \Big[(n - \frac{3}{2}) v_{r}(\theta) - v_{\theta}(\theta) \cot \theta - \frac{d v_{\theta}(\theta)}{d \theta}\Big] - v_{\theta} \frac{d \rho(\theta)}{d \theta} = 0
\end{equation}
\begin{multline} \label{vr_new}
\rho(\theta)\Big[\frac{1}{2} v_{r}^{2}(\theta) + v_{\theta}(\theta)^{2} + v_{\varphi}^{2}(\theta) - v_{\theta}(\theta)\frac{d v_{r}(\theta)}{d \theta} - 1 \Big] + (n + 1) p(\theta) \\
+ \frac{1}{8 \pi}(n - 1) b^{2}(\theta) = 0
\end{multline}
\begin{multline}\label{vt_new}
 \rho(\theta)\Big[ v_{\varphi}^{2}(\theta) \cot \theta - \frac{1}{2} v_{r}(\theta) v_{\theta}(\theta) - v_{\theta}(\theta)\frac{d v_{\theta}(\theta)}{d \theta}\Big] - \frac{d p(\theta)}{d \theta} \\
 - \frac{1}{4 \pi} b(\theta) \Big\{ b(\theta)\cot \theta + \frac{db(\theta)}{d\theta} \Big\} = 0
\end{multline}
\begin{multline}\label{vp_new}
v_{\varphi}(\theta) \bigg[ \frac{3}{2} \alpha (n - 2) p(\theta) \big(1 + \frac{b^{2}(\theta)}{8 \pi p(\theta)} \big) - \rho(\theta)\Big\{ v_{\theta}(\theta) \cot \theta
\\
+  \frac{1}{2} v_{r}(\theta)  \Big\} \bigg] - \rho(\theta) v_{\theta}(\theta)\frac{d v_{\varphi}(\theta)}{d \theta} = 0
\end{multline}
\begin{multline}\label{p_new}
p(\theta)  \Biggr\{  \gamma v_{\theta}(\theta) \frac{d \rho(\theta)}{d \theta} - (n \gamma - n - 1) v_{r}(\theta) \rho(\theta) +f (\gamma - 1)\times \\
\big(1 + \frac{b^{2}(\theta)}{8 \pi p(\theta)} \big) \bigg[ \frac{9}{4} \alpha  \rho(\theta) v_{\varphi}^{2}(\theta) \frac{\eta_{0}}{4 \pi} \Big\{ \big( b(\theta) \cot \theta + \frac{db(\theta)}{d\theta}\big)^{2} \\
+ \big(\frac{1}{2} (n - 1) b(\theta)\big)^{2} \Big\} \bigg]  \Biggr\} - \rho(\theta) v_{\theta}(\theta) \frac{d p(\theta)}{d \theta}  = 0
\end{multline}
\begin{multline}\label{d2b_new}
\eta_{0} \frac{p(\theta)}{\rho(\theta)} \Biggl\{ \Big( 1 + \frac{b^{2}(\theta)}{8 \pi p(\theta)}\Big) \Big\{ \frac{d^{2}b(\theta)}{d \theta^{2}} + \frac{db(\theta)}{d \theta} \cot\theta + \frac{1}{4}n(n - 1)b(\theta) \\
- \frac{b(\theta)}{\sin\theta^{2}} \Big\}  + \Big( b(\theta) \cot(\theta) + \frac{d b(\theta)}{d \theta} \Big) \bigg[ \Big( 1 + \frac{b^{2}(\theta)}{8 \pi p(\theta)}\Big)\Big\{ \frac{1}{p(\theta)} \frac{d p(\theta)}{d \theta}\\
- \frac{1}{\rho(\theta)} \frac{d \rho(\theta)}{d \theta} \Big\} + \frac{b^{2}(\theta)}{8 \pi p(\theta)}\Big\{ 2 \frac{1}{b(\theta)} \frac{db(\theta)}{d \theta} - \frac{1}{p(\theta)} \frac{d p(\theta)}{d \theta} \Big\} \bigg] \Biggl\} \\
+ \frac{n}{2} v_{r}(\theta) b(\theta) - v_{\theta}(\theta) \frac{db(\theta)}{d\theta} - b(\theta) \frac{dv_{\theta}(\theta)}{d\theta} = 0
\end{multline}
with six dimensionless functions $ v_{r}(\theta) $, $ v_{\theta}(\theta) $, $ v_{\varphi}(\theta) $, $ \rho(\theta) $, $ p(\theta) $ and $ b(\theta) $. This system can be solved with our boundary conditions to be express in the next section.

\section{boundary Conditions }
As we mentioned, our equation reduces to a set of six ODEs, with six dimensionless function $ v_{r}(\theta) $, $ v_{\theta}(\theta) $, $ v_{\varphi}(\theta) $, $ \rho(\theta) $, $ p(\theta) $, $ b(\theta) $, the variable $ \theta $ and six input parameters $ (\alpha, f, \gamma, n, \eta_{0}, \beta_{0}) $, where $ \beta_{0} $ is the ratio of the gas pressure to the magnetic pressure at the equatorial plane
\begin{equation}\label{b_0}
\beta_{0} = \frac{p_{gas}}{p_{mag}} \underset{90^{\circ}}{\bigg|} = 8 \pi \frac{ p}{ b^{2}} \underset{90^{\circ}}{\bigg|}
\end{equation}
which is considered to be constant. This set of OEDs can be numerically solved with proper boundary conditions. We assume the structure of the disk is symmetric to the equatorial plane, and thus we have
\begin{equation}
    \theta = 90^{\circ}: \qquad v_{\theta} = \frac{d \rho}{d \theta} = \frac{d p}{d \theta} = \frac{d v_{r}}{d \theta} = \frac{d v_{\varphi}}{d \theta} =  \frac{d b}{d \theta} = 0
\end{equation}
It is obvious that only five conditions are independent and then we require other boundary conditions. As far as we know, the mass density declines at the equatorial plane to the vertical axis. As the next condition we optimize the maximum density to unit, $ \rho(90^{\circ}) = 1 $. Now if we put the above boundary conditions into the equations (\ref{rho_new})-(\ref{d2b_new}), we obtain
\begin{equation}\label{bandary-1}
    \frac{d v_{\theta}}{d \theta}\underset{90^{\circ}}{\bigg|} = \big( n - \frac{3}{2} \big) v_{r} \underset{90^{\circ}}{|}
\end{equation}
\begin{equation}\label{bandary-2}
    \frac{1}{2} v^{2}_{r}\underset{90^{\circ}}{\big|} + v^{2}_{\varphi}\underset{90^{\circ}}{\big|}  + \bigg[ (n + 1) + \frac{(n - 1)}{\beta_{0}}\bigg] p\underset{90^{\circ}}{|} - 1 = 0
\end{equation}
\begin{equation}\label{bandary-3}
    v_{r}\underset{90^{\circ}}{|} = E_{1} p\underset{90^{\circ}}{|}
\end{equation}
\begin{equation}\label{bandary-4}
    v^{2}_{\varphi}\underset{90^{\circ}}{\big|} = \frac{E_{1}E_{3} - E_{4}}{E_{2}} p\underset{90^{\circ}}{|}
\end{equation}
\begin{equation}\label{bandary-5}
    \frac{d b^{2}}{d \theta ^{2}}\underset{90^{\circ}}{\bigg|} = \big( E_{1} E_{5} + E_{6}  \big) \frac{b}{p} \underset{90^{\circ}}{\bigg|}
\end{equation}
where
%
\begin{equation} \label{aa}
E_{1} = 3 \alpha (n - 2) (1 + \beta_{0}^{-1})
\end{equation}
\begin{equation}
E_{2} = \frac{9}{4} f \alpha (\gamma - 1) (1 + \beta_{0}^{-1})
\end{equation}
\begin{equation}
E_{3} = n \gamma - n - 1
\end{equation}
\begin{equation}
E_{4} = \frac{1}{2} f \eta_{0} (\gamma - 1) \beta_{0}^{-1} (n - 1)^{2} (1 + \beta_{0}^{-1})
\end{equation}
\begin{equation}
E_{5} = - \frac{n}{2 \eta_{0}} (1 + \beta_{0}^{-1})
\end{equation}
\begin{equation} \label{gg}
E_{6} = 1 - \frac{1}{4}n(n - 1)
\end{equation}
By substituting equations (\ref{bandary-3}) and (\ref{bandary-4}) into equation (\ref{bandary-2}), the gas pressure in the equatorial plane will be
\begin{equation}
    p\underset{90^{\circ}}{|} = \frac{-B \pm \sqrt{B^{2} - 4AC} }{2A}
\end{equation}
Here
\begin{gather}\label{ABC}
   A = \frac{E_{1}^{2}}{2}  \\
   B = \frac{E_{1}E_{3} - E_{4}}{E_{2}} + (n + 1) + (n - 1)\beta_{0}^{-1} \\
   C = -1
\end{gather}
%
\begin{figure*}
\centering
\includegraphics[width=175mm]{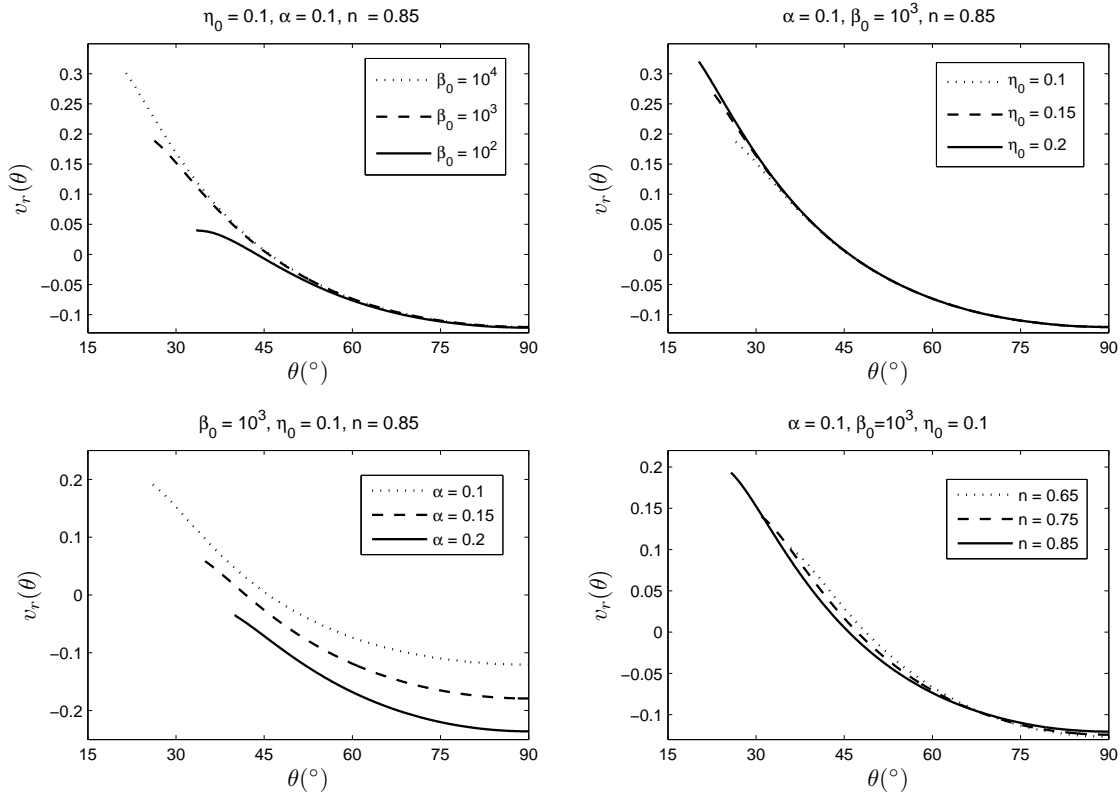}
\caption{Radial flow speed $ v_{r}(\theta) $ along $ \theta $-direction corresponding to different input parameters of $ \beta_{0} $, $ \eta_{0} $, $ \alpha $ and $ n $. Here $ \gamma = 5/3 $ and $ f = 1 $. }
\label{vrs}
\end{figure*}
%
\begin{figure*}
\centering
\includegraphics[width=175mm]{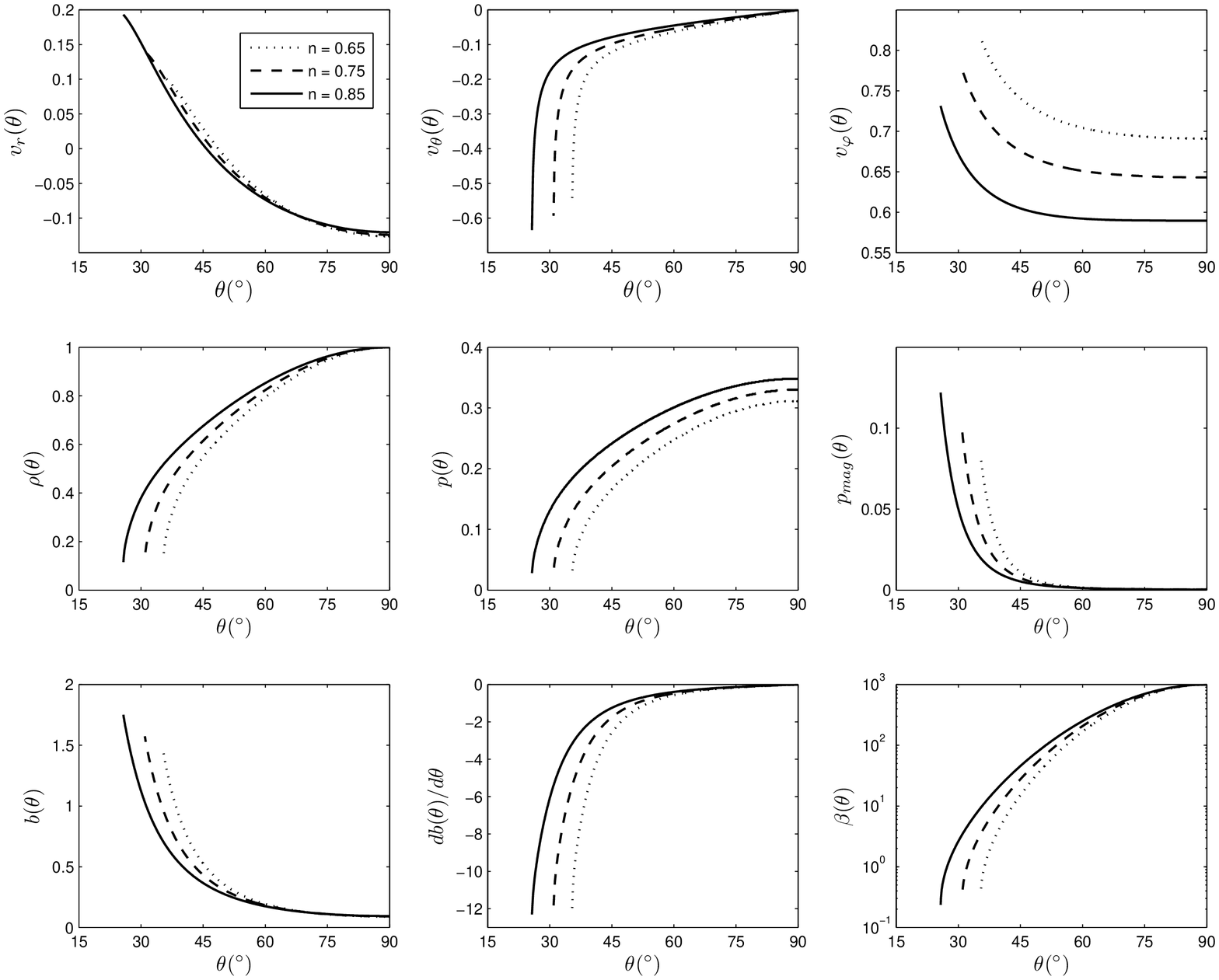}
\caption{Profiles of the physical variables corresponding to ADAFs model along $ \theta $-direction for different values of density index $ n $. The dotted, dashed and solid lines denote $ n = 0.65, 0.75 $ and $ 0.85 $ respectively. Here $ \beta_{0} = 10^{3}, \alpha = 0.1, \gamma = 5/3, f = 1 $ and $ \eta_{0} = 0.1$.}
\label{variables_n}
\end{figure*}
Having already defined the relation between gas pressure and alternative quantities, we have the value of all quantities in the equatorial plane, $ \theta = 90^{\circ} $. We know the solution with $ n = 3/2 $ examined by \citealt{ni95a}. Then following the work done by \citealt{jw}, we consider $ n \leq 3/2 $. In this way, it is seen from equations (\ref{aa})-(\ref{gg}) that $ E_{1}, E_{5} $ parameters have negative values $ (E_{1} < 0, E_{5} <0) $, while $ E_{6} $ is positive in this regime of density index. Then, it is obvious that the right hand site of equation (\ref{bandary-5}) will be positive. This indicates that at mid-plane we have $ d^{2}b/d\theta^{2} > 0 $. This very interesting mathematical result shows that magnetic pressure has a minimum value at equatorial plane. Due to this result, the ratio of gas pressure to the magnetic pressure, $ \beta(\theta) $, will have a maximum in the mid-plane of the disk. We will consider $ 10^{2} \leq \beta_{0} \leq 10^{4} $ in our figures. The upper boundary condition is adjacent to vertical axis, $ \theta = 0^{\circ} $, where the gas pressure and mass density will nearly become to zero (defining the upper limit as $ \theta_{b} $). Also there must be a $ \theta_{0} $ at a particular point where the radial velocity is equal to zero $ v_{r}(\theta_{0}) = 0 $ and after this bending point, the inflow material is deflected to outflow. Hence we can define the inflow region, where the inflow of the matter is directed toward the central mass and has a negative value, $ \theta_{0}<\theta<90^{\circ} $ and the outflow region where radial velocity has a positive value, $ \theta_{b}<\theta<\theta_{0} $. According to \citealt{xw} and \citealt{jw} the mass inflow and outflow rates can be written as
\begin{multline}\label{Mdot}
    \dot{M}(r) = \dot{M}_{inflow}(r) + \dot{M}_{outflow}(r) = 4\pi \sqrt{GM_{*}} r^{\frac{3}{2} - n} \times \\
    \bigg[ \int_{\theta_{b}}^{\theta_{0}} \rho(\theta) v_{r}(\theta) \sin d\theta + \int_{\theta_{0}}^{\pi/2} \rho(\theta) v_{r}(\theta) \sin d\theta \bigg]
\end{multline}
Based on the above equation, mass density is function of $ r $ and only when $ n = -3/2 $ the mass density becomes independent of radius. Also, if the outflow rate equals the inflow rate of material at a certain radius, the second law of thermodynamics is violated (see \citealt{xw} and \citealt{jw} for more details).\\
\begin{figure*}
\centering
\includegraphics[width=175mm]{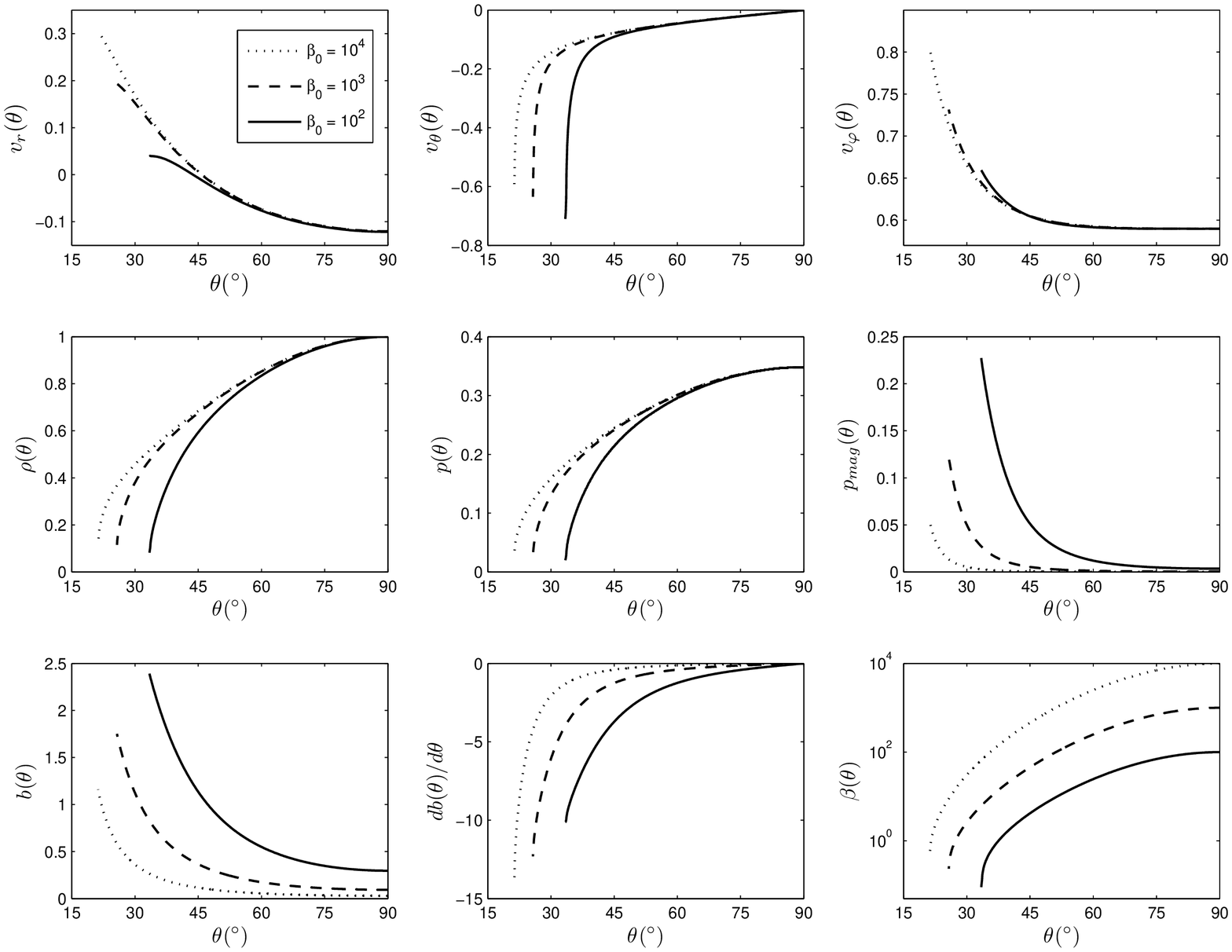}
\caption{Profiles of the physical variables corresponding to ADAFs model along $ \theta $-direction for different values of the gas pressure to the magnetic pressure in midplane of the disk,   $ \beta_{0} $. The dotted, dashed and solid lines denote $ \beta_{0} = 10^{4}, 10^{3} $ and $ 10^{2} $ respectively. Here $ \alpha = 0.1, \gamma = 5/3, f = 1, \eta_{0} = 0.1 , n = 0.85 $.}
\label{variables_beta0}
\end{figure*}
\begin{figure*}
\centering
\includegraphics[width=175mm]{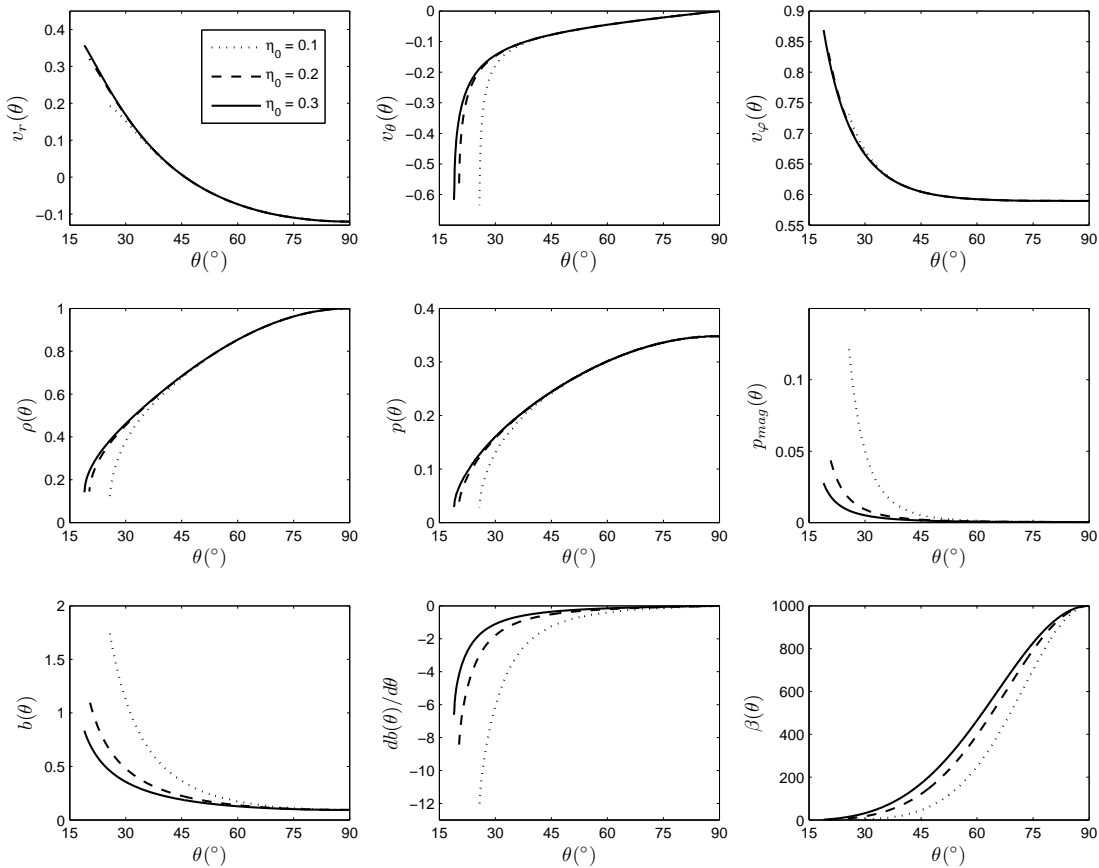}
\caption{Profiles of the physical variables corresponding to ADAFs model along $ \theta $-direction for different values of magnetic diffusivity parameter, $ \eta_{0} $. The dotted, dashed and solid lines denote $ \eta_{0} = 0.1, 0.2 $ and $ 0.3 $ respectively. Here $ \alpha = 0.1, \beta_{0} = 10^{3}, f = 1, \gamma = 5/3 $ and $ n = 0.85 $.}
\label{variables_eta0}
\end{figure*}

\begin{figure*}
\centering
\includegraphics[width=175mm]{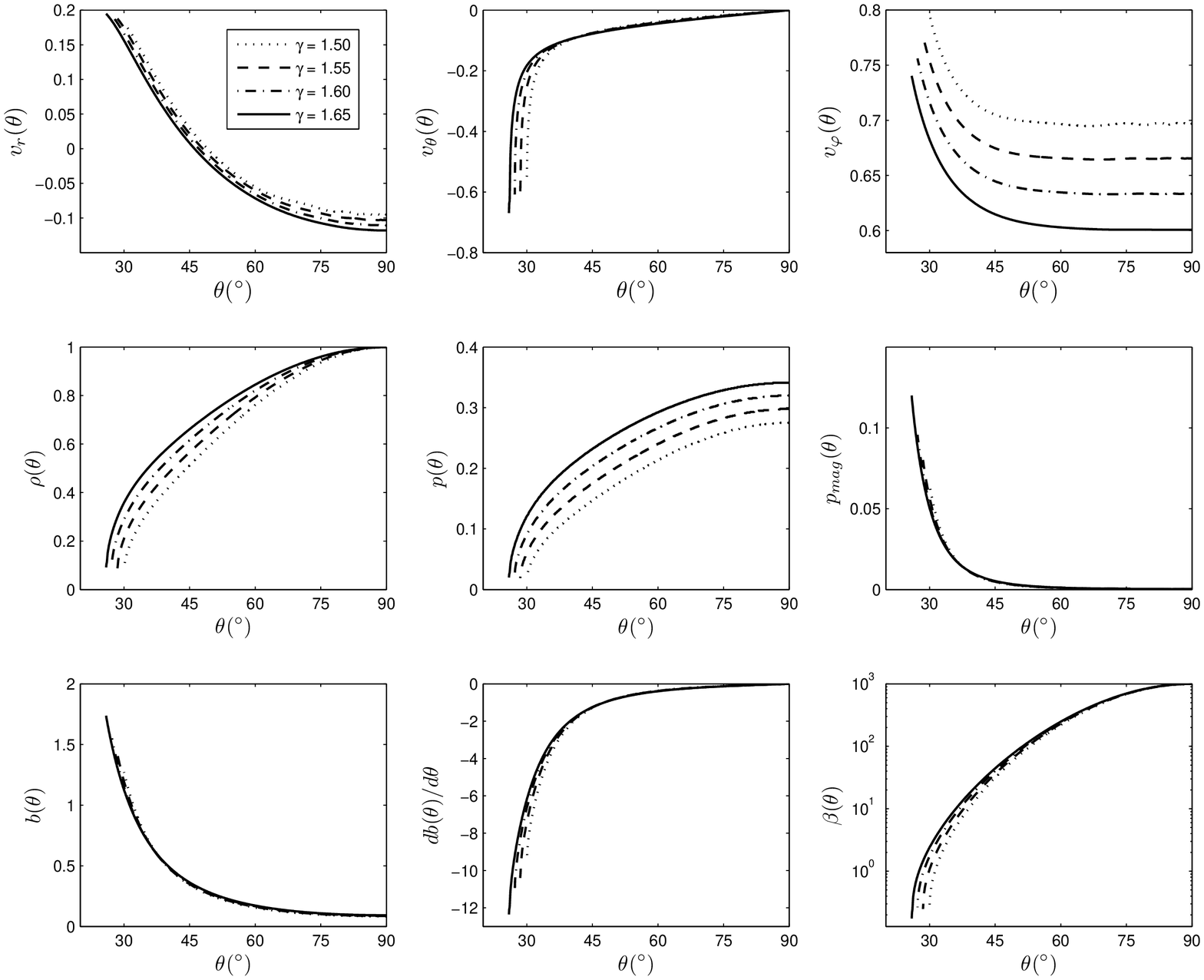}
\caption{Profiles of the physical variables corresponding to ADAFs model along $ \theta $-direction for different values of the ratio of specific heats, $ \gamma $. The dotted, dashed, dash-dotted and solid lines denote $ \gamma = 1.50, 1.55, 1.60 $ and $ 1.65 $ respectively. Here $ \alpha = 0.1, \beta_{0} = 10^{3}, f = 1, \eta_{0} = 0.1 $ and $ n = 0.85 $.}
\label{variables_gamma}
\end{figure*}

\section{Numerical Results}
In this section, we will investigate the role of magnetic field parameters $ \beta_{0} $, $ \eta_{0} $ and also $ n $, $ \alpha $ and $ \gamma $ parameters on the vertical structure of accretion disks and basic dynamical quantities of our model. As previously mentioned, we have a positive value for the radial velocity of accretion material, $ v_{r}(\theta) $. In fact this positive value indicates that the fluid is moving away from the central massive accretor and this phenomenon will escalate with the decrease of $ \theta $ and also moves towards the surface of the accretion disk. Indeed, due to the very low density around vertical axis, $ \theta = 0 $, numerical errors will appear and we have to finish our calculation at this angle. Therefore, this point is considered as the surface of the accretion disk. In other words, as is expressed in previous works (\citealt{jw}), according to the sign of the radial velocity, three different regions are distinguished in an accretion disk. One is the predominate region which is called the inflow region and starts from midplane and continues to where the radial velocity equals zero ($ v_{r}(\theta_{0}) = 0 $). Here $ \theta_{0} $ indicates the angle at which the radial velocity is zero. Actually in the interval $ \theta_{0} \leq \theta \leq 90^{\circ} $ the accretion flow moves towards the massive accretor. For this reason it is called the inflow region and it is obvious that accretion rate will be negative there. The next area that has been located just after the inflow region and continues to the surface of the disk is called the outflow region. In this area, the value of the radial velocity is positive meaning that the accretion flow is moving away from the central accretor. This region is not very wide and as it is anticipated by calculation of mass accretion rate in this area, this rate is less than unity and this does not violate the second law of thermodynamics. Also a large fraction of accretion flow at a certain radius inclines towards the massive accretor and does not escape from the disk by wind or outflow (see \citealt{xw}; \citealt{jw} for more details).  Finally, the third region is called the wind region; a very narrow area which is located between the surface of the disk and the vertical axis and we do not have many information about this area because the self-similar solution has some drawback.
\begin{figure*}
\centering
\includegraphics[width=175mm]{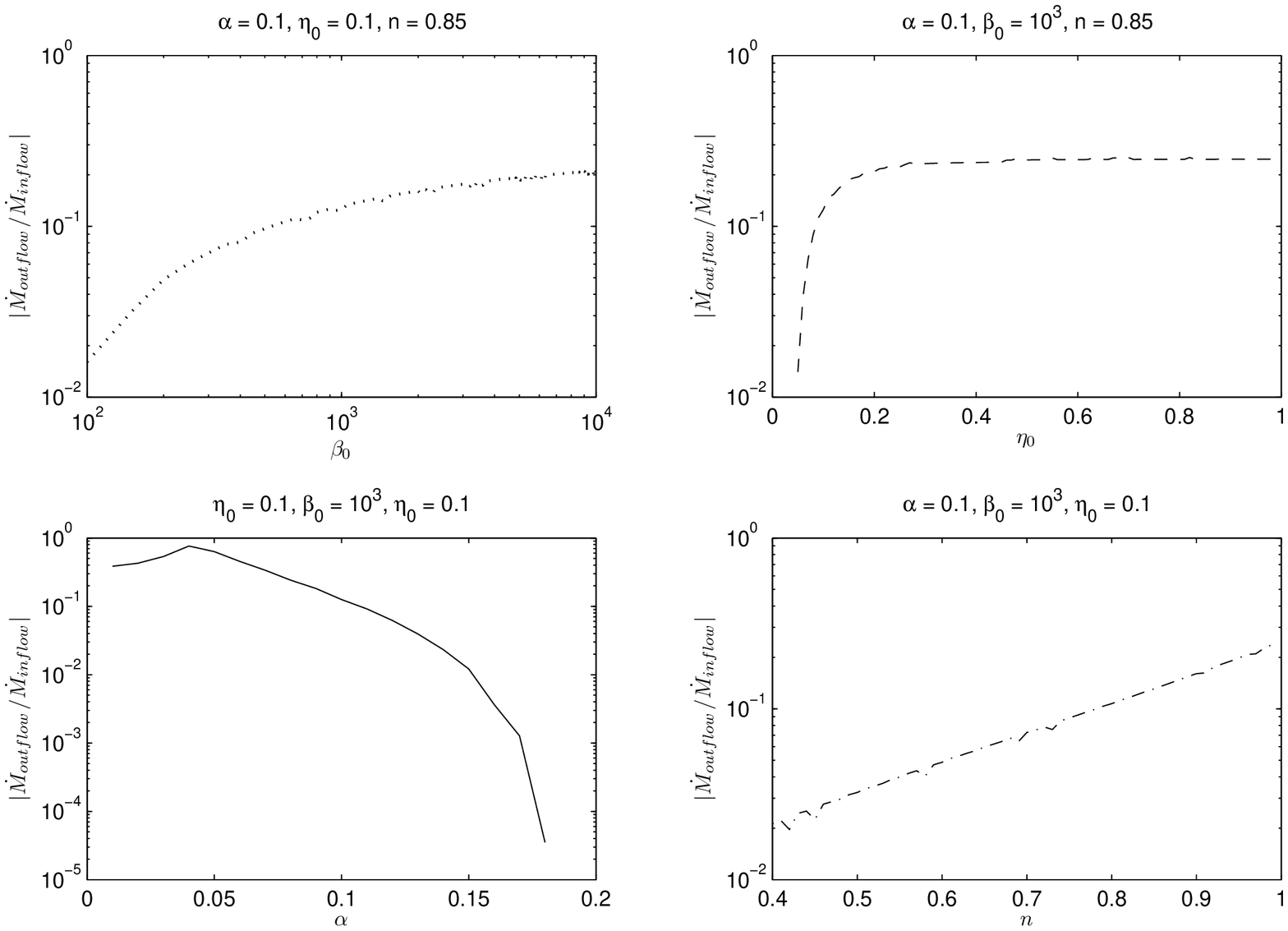}
\caption{Profiles of the radio of the outflow rate to the inflow rate corresponding to different input parameters, $ \beta_{0} $, $ \eta_{0} $, $ \alpha $ and $ n $ . Here $ \gamma = 5/3, f = 1 $.}
\label{figMdot}
\end{figure*}

Figure.~\ref{vrs}, shows the role of magnetic field parameters $ \beta_{0} $, $ \eta_{0} $, viscosity parameter $ \alpha $ and density index $ n $ on the radial velocity of the accretion flow along $ \theta $ direction. Other fixed parameters are mentioned at the top of each panel. In the top left panel, the radial velocity along $ \theta $ direction is seen with different values of $ \beta_{0} $. Here $ \beta_{0} $ demonstrates the ratio of gas to magnetic pressure in the equatorial plane. According to the Fig.~\ref{vrs}, $ \beta_{0} $ does not have any effect on the inflow region and it only changes the outflow value. It is seen that if the magnetic pressure is increased, the area of the outflow region is decreased (the small value of $ \beta_{0} $ shows high value for magnetic pressure). This implies that toroidal component of magnetic field prevents material from the surface of disk. In addition, it is clear from this panel that thickness of the disk decreases with increasing the magnetic field in the equatorial plane. The top right panel of this figure shows the role of different values of $ \eta_{0} $ on radial velocity along $ \theta $ direction. As shown in this panel, unlike the ratio of gas to magnetic pressure in midplane, the magnetic diffusivity parameter, $ \eta_{0} $, increases in the outflow region. Actually, radial velocity will be more positive in this region, but there is not any special effect on the thickness and inflow region. The bottom left panel contains different values of viscosity parameter, $ \alpha $. We assume this quantity to be between $ 0 < \alpha \leq 0.2 $ according to observations. According to this panel, the inflow area increases with increasing $ \alpha $ but the thickness of the disk and outflow region decrease. Besides, for $ \alpha > 0.2 $ the value of $ v_{r} (\theta) $ is negative. At last, the bottom right panel is plotted for different values of density index, $ n $. The anticipated value of $ n $ is smaller than unity (Yuan et al. 2012 (I, II)). This panel shows an increase in the outflow region and thickness by increasing density parameter which is in full agreement with previously studies (\citealt{xw}; \citealt{jw}). We have improved the results of the previous works by considering the toroidal component of magnetic field. Therefore, these results are very closer to reality.

In Fig.~\ref{variables_n}, we have studied the effect of mass density power-law index, $ n $ on all  physical variables of our system. As before we neglect time dependency and since we use a self-similar solution along the radial direction, all of the quantities depend on $ \theta $ only. As a result, we study the variation of quantities as a function of angle. The first row of figure. \ref{variables_n} includes components of velocity vector. In these panels, we considered $ n = 0.65 $, $ n = 0.75 $ and $ n = 0.85 $, which are shown by dotted, dashed and solid line respectively. We can see that the radial velocity increases as density index enhances but other components of velocity vector $ v_{\theta}(\theta) $ and $ v_{\varphi}(\theta) $ decrease. According to our symmetry the value of $ v_{\theta}(\theta) $ on the equatorial plane is zero, i.e. $ v_{\theta}(90^{\circ}) = 0 $. Also the variations of the second component of velocity vector is nonzero, $ d v_{\theta}(90^{\circ})/d \theta \ne 0 $. As a result of this symmetry, the fluid is just directed to the central object in midplane and then by moving towards the vertical axis we can see that the direction of material becomes inverse and is inclined towards the surface of the disk. This is because the radial velocity of accretion flow is positive which has been considered in previous works (\citealt{xw}; \citealt{jw}). An important issue is sup-Keplerian velocity. In fact all three components of velocity in all angles have smaller values than Keplerian. The second row of Fig. \ref{variables_n}, shows the variations of mass density, gas pressure and magnetic pressure of fluid by different values of density parameter, $ n $. Since the starting point is on the equatorial of the disk and the maximum value of density belongs to this point, by gradual motion of the surface of the disk, we can show a decrease in density panel, which appears in each three parameters of $ n $. Also similarly, the gas pressure will decrease as $ \theta $ declines. The interesting and new behavior are related to magnetic pressure variations based on $ \theta $. As shown in this panel magnetic pressure increases as the angle decreases by motion toward surface of disk. Actually by taking this symmetry assumption on the equatorial plane of the disk and the positive sign for second derivation of magnetic field at this point, $ d B_{\varphi}(\theta) / d \theta $, it is due to the fact that the minimum value for magnetic field is in the equatorial plane, meaning that an increase of field as $ \theta $ falling down is not far from reality . As a result, the magnetic field on the disk surface is an important parameter for determining the behavior and structure of the disk. Additionally, as shown in these panels, this parameter leads to an increase in the thickness. The last row of Fig.~\ref{variables_n} is related to the magnetic field quantities and also the ratio of gas pressure to magnetic pressure, $ \beta(\theta) $. It shows that $ b(\theta) $, (the toroidal field parameter) increases but derivatives of magnetic field, $ d B_{\varphi}(\theta) / d \theta $ and $ \beta(\theta) $ decrease as $ \theta $ decreases. According to \citealt{jw}, if the radial components of pressure gradient, gravity force,  magnetic force  and centrifugal force are caused by going up or down in the radial velocity according to the terms  in equation (\ref{vr_new}) and also if we assume a value of density index  $ n < 1 $, the magnetic force term will be negative and it can be a justification for decrease of outflow velocity value in proportion to figures in \citealt{jw} (see figures 3 \& 6 in \citealt{jw} for more details).

Figure. \ref{variables_beta0} shows the effect of various values of magnetic field parameter $ \beta_{0} $ on the profiles of the physical variables. As it is stated before, $ \beta_{0} $ is the ratio of gas to the magnetic pressure in the midplane of the disk. We adopt $ \alpha = 0.1 $, $ \gamma = 5/3 $, $ f = 1 $, $ \eta = 0.1 $ and $ n = 0.85 $. In this figure, dotted, dashed and solid line correspond to $ \beta_{0} = 10^{4} $, $ \beta_{0} = 10^{3} $ and $ \beta_{0} = 10^{2} $ respectively. The profiles of velocity components are shown in the top row of this figure. we see that by rising in magnetic parameter $ \beta_{0} $ in the equatorial plane of the disk, all three components of velocity vector decrease toward surface (notice that the small value of $ \beta_{0} $ present stronger magnetic filed in midplane of the disk). Then we deduct that velocity filed reduce by enhancing in magnetic parameter $ \beta_{0} $. Also the thickness of the disk will decrease too. Moreover according to middle profiles of figure. \ref{variables_beta0} it is predicted that the mass density and gas pressure decrease along $ \theta $ direction toward surface of the disk. On the other hand, the magnetic pressure enhance and become strong near surface of the disk by increasing of $ \beta_{0} $. The identical behavior is shown in profile of $ b(\theta) $ in the below row of figure. \ref{variables_beta0}. Moreover when we move toward surface the variation of magnetic field will reduce. As the initial condition set in equatorial plane and the value of this variable is considered equal zero there. Then we show a negative value for variation of magnetic filed all over the direction. As is is known the behaviors of gas and magnetic pressure then the ratio of these variable, i.e., $ \beta(\theta) $, reduce along $ \theta $ direction from equatorial plane to the surface of the disk.

One of the other important input parameters in our model is  magnetic diffusivity parameter, $ \eta_{0} $, and the major behavior has been shown in figure. \ref{variables_eta0}. The dotted, dashed and solid lines correspond to $ \eta_{0} = 0.1 $, $ \eta_{0} = 0.2 $ and $ \eta_{0} = 0.3 $ respectively. We supposed that $ \alpha = 0.1 $, $ \beta_{0} = 10^{3} $, $ f = 1 $, $ \gamma = 5/3 $ and $ n = 0.85 $. The radial velocity and angular velocity will increase as $ \eta_{0} $ increases. While $ v_{\theta}(\theta) $ component will be decreased. It also increases the parameter $ \eta_{0} $ which will be caused by a significant enhance in thickness of ADAFs. According to figure. \ref{variables_eta0}, we can see that magnetic pressure decrease with $ \eta_{0} $. However, we are seeing an increase in magnetic pressure as angle decreases and also moving towards disk surface and approaching the vertical axis. The last row of figure. \ref{variables_eta0} is dedicated to variations of the magnetic field and also ratio of gas pressure to magnetic pressure. By looking at these figures we can conclude that $ \eta_{0} $ parameter does not have any shift on inflow region also it change the outflow area. As a result due to the symmetry considered in equatorial plane previously mentioned we can see an increase in the magnetic field from equatorial plane to the disk surface. The variation of magnetic field will increase by increasing $ \eta_{0} $. Moreover, a decrease in the quantity $ \beta(\theta) $ can be seen by decreasing angle which reflects the fact that while moving towards the disk gas pressure is reduced and magnetic pressure will increase and the result of this process of reducing in the value of $ \beta(\theta) $  will be in motion to vertical axis.

Another input parameter of the model is $ \gamma $ that has been shown in figure . \ref{variables_gamma}. We assume that $ \alpha = 0.1 $, $ \beta_{0} = 10^{3} $, $ f = 1 $, $ \eta_{0}= 0.1 $ and $ n = 0.85 $. As shown the figure. \ref{variables_gamma} the increase of thickness by increasing $ \gamma $ parameter. The initial value of radial velocity at the equatorial plane reduces with $ \gamma $ and actually it will be much more negative than before which represents the inflow region will be increases. Moreover this parameter will make the  $ v_{\theta}(\theta) $ component  of velocity negative and also will have a significant reduction in the rotational speed of the accretion flows. According to the diagram the magnetic pressure and the magnetic field are observed these quantities with increasing parameter $ \gamma $ will be increases in the moving disk surface increases. The panels of  variations of field and the ratio of pressures will decreases as $ \gamma $ decreases.

On principle this value must be smaller than the unity in order to not violate the second law of thermodynamics.  Figure. \ref{figMdot} shows fundamental variations in the quantity of discs with the parameters $ \beta_{0}$, $\eta_{0}$, $\alpha$ and $ n $ in four separate panels. Fixed parameters corresponding to each of the graphs are listed at the top of each ones. The top left panel shows the ratio of outflow accretion over inflow accretion depending on different values of $\eta_{0}$ between $ 10^{2} < \beta_{0} < 10^{4} $. As previously mentioned if $ \beta_{0} $ decreases it shows the strong effect of magnetic field at the start point i.e., in the equatorial plate. Then according this panel it can be argued reduce the magnetic field is increased, the proportion. Also this rate is smaller than unit which expresses most of the material increases toward to accretor and only a fraction of that moves out of disk , finally out be a form of wind and jet. In the top right panel the ratio of accretion rates to $ \eta_{0}$ is also shown. A very interesting point in this diagram is that the $ \eta_{0} $ parameter only increases at outflow boundary and does not have a significant effect on inflow. Now from this information it can be inferred that increasing the value of $\eta_{0}$ in the interval between $ 0 < \eta_{0} < 0.2 $ has been dramatic and the interval  $ 0.2 < \eta_{0} < 1 $ behavior of charts are fixed. The bottom left panel also shows the behavior of mass accretion rates to variation of viscosity parameter, $ \alpha $. We also considered the interval $ 0 <  \alpha < 0.2 $ for $ \alpha $, because it is in adaptation with observations. We see a double behavior of viscosity parameter. In fact, at first we see the proportion of outflow ratio to inflow ratio. Around the $ \alpha = 0.05 $ this increasing trend has stopped and the $ 0.05 < \alpha < 0.2 $ interval reduction in the diagram will be done. As a result we will see the peak of the figure near $ \alpha = 0.05 $. This effect has also been seen by  \citealt{jw} paper about ADAFs, figure 17,  and for more information you can be refer to this article. Finally, the bottom right of Figure. \ref{figMdot} shows the behavior of ratio mass accretion rates to variation of mass density power-law index, $ n $. This can be understood by referring to previous figures because we are seeing an increase in the ratio of outflow rate to inflow rate along $ n $, then mass density parameter, $ n $,  will be increase the outflow boundary that is in perfect harmony by Fig \ref{variables_beta0}. According to the last simulation works, we have considered smaller range than previous one, i.e., $  0 < n < 1$. This increasing trend has also been observed in Figure 17 of \citealt{jw}.
\section{discussion}
The main aim of this manuscript is verifying the structure of advection dominated accretion flows $(ADAFs)$ along the $\theta$ direction when bathed in a toroidal magnetic field. The results have shown that the vertical structure of the disks is significantly affected by the magnetic field and its correspond resistivity. By the self-similar solution along radial direction, the proper boundary conditions and reflection symmetry in equatorial plane of the disk, we have constructed the structure of the disk along the $\theta$ direction explicitly. We have shown that only by assuming $v_{\theta}\ne 0$ the solutions represent a inflow-outflow behavior which is not reported in the pervious ADAFs investigations. This assumption improved \citealt{ni95a} solution to interpret the existence of outflow in the hot accreting systems. Our disk consists of three different regions: 1- predominate region which is called the inflow region. It start up from midplane to where the radial velocity is equal zero, $ v_{r}(\theta_{0}) = 0 $ and contains the largest portion of mass. 2- an outflow region which located just after the inflow region and will continue to the surface of the disk in which the matter starts escaping the central accretor in the $r$-direction. 3- the third region, called wind region. This region contains the material blowing out from the boundary of the outflow region. An area with very low wide and since the self-similar solution has some drawback, we do not have many information about it. Compared to the nonmagnetic field solution \citealt{jw}, the existence of magnetic field and it's resistivity in our case, can produce more advective energy. Also, the B-field configuration can affect the energy transportation along accretion disks.

In this paper we used two parameters $ \beta_{0}$, $\eta_{0}$ in order to study the effect of magnetic field on the vertical structure of the disk. Our results shown that the magnetic pressure will increase if $\theta$ changes moving towards the surface of disc. Also if we increase the $ \beta_{0}$ parameter (which represents the ratio of gas pressure to field pressure at $ \theta =  90^{\circ}$) it will enhance the rate of magnetic pressure growth (figure 3). Moreover, it was demonstrated that the disc's half width declines as $ \beta_{0}$ increases (figure 3).

In a real accretion disk, there are several important processes other than viscosity and resistivity which they are also expected to have influenced on the dynamical structure of the disks such as radiation pressure, photon trapping and any types of instabilities. It is immediately clear that the changes in the boundary conditions affect the structure of the solutions and also it should be emphasized that we obtained this solution in a steady state regime. Thus, the mere existence of a self-similar solution in no way guarantees that the solution is relevant to real accretion flows. Given these facts, the treatments in the paper are very simplified, however sufficiently general to describe many disk/outflow systems. Although we have made some simplifications in order to treat the problem analytically, the presented self-similar solutions shown that the input parameters can really change the typical behavior of the physical quantities of ADAF disks. Besides, numerical simulations support that the mass accretion rate, radial velocity and density can be well approximated as power-laws (see, e.g. Stone, Pringle \& Begelman 1999; Yuan, Bu \& Wu 2012). All these power-law profiles justify the self-similar methodology adopted in our work. Although these self-similar solutions are too simple to make any comparison with observations, however, we think that one may relax the self-similarity assumptions and solve the equations of the model numerically. This kind of similarity solution could greatly facilitate testing and interpretation of the results.

We thank Fu-Guo Xie and Feng Yuan for their discussions and useful suggestions. We are also like to appreciate the referees for their thoughtful and constructive comments which clarify some points in the early version of the paper.

\end{document}